\documentstyle[12pt,aasms4]{article}

\slugcomment{submitted to the Ap.J.}

\lefthead{Della Valle \& Panagia}
\righthead{SNIa in Radiogalaxies}

\newcommand{\ie}{\emph{i.e.}\ }
\newcommand{\etal}{\emph{et al.}\ }
\newcommand{\eg}{\emph{e.g.}\ }
\newcommand{\cf}{\emph{c.f.}\ }

\newcommand{\msun}{M$_\odot$}
\newcommand{\kms}{km~s$^{-1}$}

\newcommand{\lsim}{{\, \lower2truept\hbox{
${< \atop\hbox{\raise4truept\hbox{$\sim$}}}$}\,}}
\newcommand{\gsim}{{\, \lower2truept\hbox{
${> \atop\hbox{\raise4truept\hbox{$\sim$}}}$}\,}}

\begin{document}

\title{The rate and the origin of type Ia supernovae in radiogalaxies}
\bigskip

\centerline{Massimo Della Valle\footnote{Istituto Nazionale di
Astrofisica, Osservatorio Astrofisico di Arcetri,\\ Via Enrico Fermi 5,
I-50125 Florence, Italy; {\it massimo@arcetri.astro.it}} and Nino
Panagia\footnote{Space Telescope Science Institute, 3700 San Martin
Drive, Baltimore, MD~21218. On assignment from the Space Telescope
Operations Division, Research and Scientific Support Department of ESA;
{\it panagia@stsci.edu}}}

\begin{abstract}

An analysis of type Ia supernova (SNIa) events in early type galaxies
from Evans \etal~(1989) database provides strong evidence that the
rate of type Ia supernovae (SNe) in radio-loud galaxies is about 6
times higher than the rate measured in radio-quiet galaxies, \ie~
SNIa-rate$(radio-loud~galaxies)=0.47^{+0.23}_{-0.15}~h^2_{50}$ SNe per
century and per 10$^{10}$L$^{B}_\odot$ (SNU) as compared to
SNIa-rate$(radio-quiet~galaxies)\lsim 0.080~h^2_{50}$ SNU. The exact
value of the enhancement is still rather uncertain, but is likely to
be in the range $\sim 4-15$. We discuss the possible causes of this
result and we conclude that that the enhancement of SNIa explosion
rate in radio-loud galaxies has the {\it same common} origin as their
being strong radio sources, but that there is no causality link
between the two phenomena.  We argue that repeated episodes of
interaction and/or mergers of early type galaxies with dwarf
companions are responsible for inducing both strong radio activity in
$\sim$14\% of early type galaxies, and the $\sim1$~Gyr old stellar
population needed to supply an adequate number SNIa progenitors.
Within this scenario we predict that the probability of detecting a
core-collapse SN event in radio-loud elliptical galaxies amounts to
about 4\% of their SNIa events.

\end{abstract}



\section{Introduction}

Supernovae (SNe) are milestones of galaxy evolution, and the knowledge
of the supernova rate is an essential means for constraining the
mechanisms of galaxy formation and for understanding galaxy evolution.
For example, Canizares, Fabbiano \& Trincheri (1987) find that the
explosion of type Ia SNe can have a strong influence on the properties
of the X-ray emitting gas in early type galaxies, and Renzini \etal~
(1993) argue that the production of iron in systems that at the
present epoch have a reduced or totally exhausted star formation
activity, such as the elliptical galaxies, dramatically depends on the
rate of type Ia events. Thus, any mechanism capable of producing an
enhancement of SNIa rate in this type of galaxies should be regarded
with the highest interest. Two recent works, by Capetti (2002) and by
Livio, Riess and Sparks (2002), deal with this possibility. Capetti
(2002) finds a close association between the position of 6 type Ia SNe
discovered in radiogalaxies and their jets, while Livio \etal~(2002)
claim the detection of an enhancement of the nova rate in M87 in the
regions near its jet. The former result would indicate that type Ia
SNe in radiogalaxies show a statistically significant preference (at
the level of $\sim 95\%$) to occur close to their jets. The latter
result may suggest that the presence of jets could increase the
accretion rate from the interstellar medium onto the white dwarfs up
to $\sim 10^{-9}$M$_\odot$ yr$^{-1}$, which is the rate necessary to
trigger classical nova explosions (Prialnik \& Kovetz 1995). By
combining the two results, one would infer that the presence of the
jets may increase the efficiency of accretion rate from the ISM and
drive an accreting white dwarf to approach the Chandrasekhar mass and
ignite, according to the standard single-degenerate model, a SNIa
explosion (Livio \etal 2002, and references therein). In this scenario
it is the jet itself that alters the physical conditions of its
environment and thus it is the {\it direct cause} of a local
enhancement of nova and type Ia SN explosions.

Since jets are outstanding signatures of radiogalaxies (e.g. Bicknell
2002), it is important to extend the investigation to all galaxies with
strong radio emission and to study the SNIa rate as a function of the
radio activity. In this Letter we will use the available SN databases
to investigate if, and to which extent radiogalaxies exhibit an
over-production of type Ia SNe with respect to quiescent galaxies of
the same class of luminosity, and to clarify what is the dominant
mechanism that drives this process. In the following, we will denote as
{\sl radio-loud} sources, or {\sl radiogalaxies}, all galaxies which
have radio luminosities at 1.4GHz higher than $10^{29}$ erg s$^{-1}$
Hz$^{-1}$, and M$_R < -20.5$ (e.g. Sadler, Jenkins, \& Kotanyi 1989,
Ledlow \& Owen 1996) and {\sl radio-quiet} all galaxies with radio
luminosities below that threshold.  The threshold value of $\sim
10^{29}$ erg $s^{-1}$ Hz$^{-1}$ is the bottom limit of the luminosity
function for the radio-galaxies of `low' radio luminosity  (see also
Urry \& Padovani 1995), and, therefore, it represents the
most natural threshold to identify  ``radio-loud'' galaxies.


\section{Statistics of SNIa explosions in early type galaxies}


Among the several databases on supernovae currently available, we 
consider the one compiled by van den Bergh, Clure and Evans (1987) and
Evans, van den Bergh and Clure (1989), which has the advantage of being
based on a very homogeneous and systematic set of observations so as to
provide, in addition to other relevant parameters, the control time for
each monitored galaxy, which is a fundamental quantity to correctly
estimate any SN explosion rate. In addition we note that the Evans's
sample of galaxies was necessarily defined `a priori', i.e. the
galaxies to be monitored were selected before the SN explosions, and
therefore our statistical  analysis doesn't suffer of any kind of `a
posteriori' selection criteria, which actually may affect the
statistical analysis carried out on compilations of parent galaxies of
SNe.

\begin{table}
\begin{center}
\caption{Surveillance time (yr$\times$L$^B/10^{10}$L$_\odot^B$) and
number of SNIa events as a function of M$_B$}
\label{time}
\begin{tabular}{lrrrrrrr}
\hline
galaxies   & $N$ & M$_B=-19$ & M$_B=-20$  & M$_B=-21$ & M$_B=-22$ & M$_B=-23$ & Total~~~ \\
          &     &           &            &           &           &
          &yr $10^{10}$L$_\odot^B$\\
\hline
radio-quiet   & 178 &  270 (0)~~ &  627 (0)~~ &  1274 (0)~~ &	99 (0)~~ &    0 (0)~~ & 2270 (0)~~  \\
radio-loud    &  19 &    0 (0)~~ &   43 (0)~~ &   221 (1)~~ &  378 (1)~~ &  205 (2)~~ &  847 (4)~~  \\
total         & 197 &  270 (0)~~ &  670 (0)~~ &  1595 (1)~~ &  477 (1)~~ &  205 (2)~~ & 3117 (4)~~  \\
\hline
\end{tabular}

\end{center}
\end{table}

Evans \etal~(1989) report 24 supernovae discovered during the period
1980-1988, 4 of which, SN 1980N, SN 1981D, SN 1983G and SN 1986G, are
type Ia discovered in early type galaxies [(in NGC 1316 (2), in NGC
4753 (1) and NGC 5128 (1)]. Fig. 1 shows the frequency distribution of
197 early type galaxies (from E to S0/a) monitored by Evans \etal~1989
(their tab. 2) as a function of their absolute B-band magnitude,
M$_B$.  To be consistent with Evans \etal~(1989), the absolute
magnitudes of the galaxies have taken from the Revised Shapley-Ames
Catalog (Sandage \& Tammann 1981) in which $H_0=50~km~s^{-1}~Mpc^{-1}$
is adopted. In tab.~1 we report the galaxy type (col. 1), the number of
galaxies for each type (col. 2), the normalized control times in years
for bins of one magnitude for radio-quiet and radio-loud galaxies
(col.  3,4,5,6,7), the number of type Ia SNe discovered (in brackets)
and the total surveillance time (col.8). Following Evans \etal the
surveillance time for each galaxy is the product of the control time,
\ie the time during which a given galaxy has been observed, to the
galaxy's B-band absolute luminosity, so that is given in units of years
$\times$ 10$^{10}$ L$^B_\odot$. This indeed is the appropriate quantity
to use in order to derive SN explosion rates per unit luminosity.
An inspection to tab. 1 reveals the following facts:

i) Evans' sample is not biased in favor of bright galaxies.  As
apparent from Fig.~1, the distribution of monitored galaxies agrees
remarkably well with a typical luminosity function of early type
galaxies, i.e. a Schechter's (1976) function with $\alpha=0.2$ and
M$^*$=--20.88 for an adopted H$_0$=50~\kms~Mpc$^{-1}$ (\eg Muriel,
Nicotra \& Lambas 1995).

ii) The fraction of radio galaxies in Evans' sample is  
$0.096\pm0.022$ of the whole sample of galaxies (\cf Fig.~1).  Such
fraction is marginally smaller than $0.140\pm0.024$ as reported by Ledlow
\& Owen (1996) for radio galaxies having the same radio luminosities
($>10^{29}$ erg s$^{-1}$ Hz$^{-1}$ at 1.4GHz). This  indicates that the
frequency of radio galaxies in Evans' sample is not affected by biases
related to their radio properties.

iii) Considering all early type galaxies, regardless of their radio
properties, an overall SNIa rate of $0.13^{+0.06}_{-0.04}~h^{2}_{50}$
SNU\footnote{$h^{2}_{50}= H_0/$(50~\kms~Mpc)$^{-1}$; $1SNU=1
SN(100yr)^{-1}$ (10$^{10}L^B_\odot$)$^{-1}$} is obtained
\footnote{Differences with the  rates listed by Evans \etal (1989) are
due to obvious typos in their paper. It is  easy to verify that the
data reported in their Tab. 4 (col. 2) are inconsistent with the rates
of  Tab. 6 (col. 2 and 5).}. This value compares favorably with the
most recent determination of the rate of type Ia SNe in early type
galaxies by Cappellaro \etal~(1999) who give an average of $0.080\pm
0.027$ SNU (for an adopted H$_ \circ=50$ km s$^{-1}$ Mpc$^{-1}$), and
confirms the high statistical value of Evans' sample.

iv) Separating radio-loud from radio-quiet galaxies, we have 4 SNIa
events in radio-loud galaxies and 0 in radio-quiet galaxies.  
Therefore, we can estimate the  rate of type Ia SNe for
radio-loud galaxies to be  $0.47^{+0.23}_{-0.15}~h^{2}_{50}$ SNU (the
errors are estimated from a direct application of Poisson statistics).
The SNIa rate in radio-loud galaxies is a factor $\sim6$ higher than
the one deduced  by Cappellaro \etal~(1999) for the overall rate of
SNIa events in early type galaxies.  This is clear evidence  for a
strong enhancement  of SNIa explosions in radio galaxies, especially
considering that Cappellaro \etal~rate includes both contributions from
quiescent and radio galaxies, and, therefore, the actual ratio of SNIa
rates in radio-loud galaxies to radio-quiet galaxies should be
appreciably higher than $\sim6$, say, 10 or more.

The null result for radio-quiet galaxies can also provide interesting
constraints on the SNIa rate in theses galaxies.  In particular, from
simple Poisson statistics we find that the probabilities of obtaining a
null result are 50\%, 16\% and 5\% when the  expected values are 0.7,
1.8, and 3.0. Correspondingly, one should expect that the SNIa rate in 
radio-quiet galaxies be likely higher than the 50\% probability value,
\ie~0.03$h^{2}_{50}$~SNU, possibly around the 16\% value, 
0.08$h^{2}_{50}$~SNU,  and hardly higher than the 5\% probability
value, 0.13$h^{2}_{50}$~SNU.  Note that since 0.13$h^{2}_{50}$~SNU is
equal to the estimated overall SNIa rate, it must really regarded as a
strict upper limit to the SNIa rate for radio-quite galaxies. 
Therefore, just using results from Evans' sample, we conclude that the
ratio of the SNIa rate in radio-loud galaxies is definitely higher than
0.47/0.13=3.6 times the one in radio-quiet galaxies, and may be as high
as 0.47/0.03$\simeq$15.


\section{Discussion and Conclusions}

An analysis of Evans \etal~(1989) database provides strong evidence
that the rate of type Ia SNe in radiogalaxies is about 6 times higher
than the rate measured in quiescent galaxies, \ie~
$0.47^{+0.23}_{-0.15}~h^{2}_{50}$ SNU (this paper) as compared to
$\lsim 0.080\pm 0.027$ $h^{2}_{50}$SNU as deduced approximately from
Evans' sample itself, or inferred from Cappellaro \etal~(1999) results
based on the analysis of a richer sample of early type galaxies. The
actual value of the enhancement is still uncertain, due to the
use of somewhat scanty statistics, but is likely to be in the range
$\sim 4-15$. Databases of SNe currently available in literature and
characterized by larger number of SNe than the Evans one, do not
report the complete set of data, \ie the sample of monitored galaxies
and the control time for each of them, and, therefore, they cannot be
used for our detailed analysis. Once the correlation of a high SNIa
rate with strong radio emission is established, one wonders what is
the causality link between the two phenomena.  {\it A priori}, the
strong enhancement of the rate of SNIa events in radio galaxies may be
explained in basically three different ways:

1) {\it The high SNIa explosion rate is the cause of strong radio
emission in some elliptical galaxies.}  This scenario is as
interesting as unlikely.  First of all, it is generally accepted that 
that the bulk of the radio emission from strong radio galaxies is
mostly associated with relativistic jet activity, presumably powered by
accretion on central massive black holes.  Second, one has to consider
that substantial  radio emission is associated exclusively with  core
collapse SNe (at a level of several $10^{27}$ erg $s^{-1}$ Hz$^{-1}$;
see, e.g. Weiler et al. 2002), but no type II or Ib/c SN
has ever been discovered in early type galaxies (van den Bergh et al.
2002). Under the hypothesis that the whole core-collapse SN population
be responsable for the observed radio emission of radiogalaxies, one
should invoke the simultaneous presence of several hundreds to many
thousands core collapse SNe in radio galaxies and require that they all
be heavily obscured.  This possibility is very unlikely because, among
other problems, would require a SN rate of about $10^4$ SNe/century or
higher: this is a very high value that is inconsistent with
observational evidence. Another possibility to consider is that it is
the kinetic energy from SN explosions that feeds the strong emission of
radio-loud ellipticals. However, the absolute rate of SNe of all types 
{\it per unit mass} in elliptical galaxies, even with strong radio
emission, is more than 10 times  lower than that of late type galaxies
(\eg Della Valle \& Livio 1994, Panagia 2000,  Mannucci \etal 2003).
Therefore,  the input of kinetic energy from SN explosions is at least 
10 times lower  in radio loud ellipticals than it is in normal spirals,
but,   at the same time, the radio power from spirals is substantially
lower than that of radio-loud elliptical galaxies (\eg Gioia, Gregorini
\& Vettolani 1981).  Since the ISM  conditions greatly  favor  spiral
galaxies over ellipticals in transforming the SN kinetic energy  into
radio emission, and, still, the former are weaker radio emitters than
the latter, we can confidently exclude that SNe be the cause of strong
radio emission in elliptical galaxies.

2) {\it The enhanced SNIa explosion rate in radio-loud galaxies is a 
direct consequence of their being strong radio sources}.  This is the
scenario suggested by Capetti (2002) and Livio \etal~2002. In
radiogalaxies the accretion rate from the interstellar medium onto
white dwarfs could be enhanced by the action of the jets. This process
could also account for the high rate of novae observed in M87 in the
regions nearby the jets. The association between type Ia SNe and jets
requires a further step, \ie~that the progenitors of these SNeI-a are
cataclysmic-type systems, according to the single degenerate scenario.
In this hypothesis, the enhancement of SNIa in radio galaxies is
expected to be spatially confined to the regions immediately adjacent
to radio jets and/or the bulk of radio activity.  While this may be
true for Virgo~A=M87, it appears not to be the case for
Fornax~A=NGC~1316, in which the two SNIa (SN 1980N and SN 1981D) are
located quite far from strong radio lobes (Geldzahler \& Fomalont
1984), nor for Centaurus~A=NGC~5128, in which SN~1986G is deeply
embedded in the equatorial lane of gas and dust. At any rate, if the
dominant phenomenon is accretion from the ISM rather than from a binary
companion, an extreme consequence would be that {\it single} WDs will
contribute to Nova and perhaps SNIa production, and not only binary
systems with suitable parameters.  Actually, since the fraction of
binary systems that may give rise to a SNIa is only 5-10\% of the total
number of systems with masses above 3 \msun~(Madau, Della Valle \&
Panagia~1998), one should expect a very dramatic increase of the SNIa
rate in jet dominated radio-galaxies, by factors as high as 30 or
higher if direct accretion from ISM becomes the dominant accretion
process.

3) A third possibility that we strongly favor is that {\it the
enhancement  of SNIa explosion rate in radio-loud galaxies has the same
common origin as their being strong radio sources, but there is no
causality link between the two phenomena}.  Indeed, the radio activity
of a galaxy is most likely triggered by interaction and/or mergers
(e.g. Baade \& Minkowski 1954, Balick \& Heckman 1982, Heckman
\etal~1986). Also, the SNIa rate can be enhanced by the formation or
the capture of relatively young stellar populations in which SNIa occur
at much higher rates than in genuinely old populations. SNIa
progenitors are stars with original masses above $\sim$3~\msun, having
nuclear-burning lifetimes shorter than $\sim$400~Myrs. After that time,
the star becomes a white dwarf and may explode as a SNIa only after an
additional time as needed for either accreting mass from a companion or
merging with it. Such a time is essentially unconstrained by theory,
but measurements of SNIa rates at different redshifts suggest that it
should be shorter than $\sim$1~Gyrs (\eg~Madau \etal~1998, Pain \etal
2002).  As a consequence, in order to sustain a substantial rate of
SNIa explosion, a galaxy has to have a steady supply of young stellar
populations, at least over a time-scale of 1-2~Gyrs. In late type
galaxies, the active star formation provides the needed input for a
steady production of SNIa events. Such a supply in early type galaxies
is naturally provided by repeated episodes of interaction or mergers
that induces either formation of young stellar populations (galaxy
interaction; \eg the Antennae galaxies, Whitmore \etal~1999) or capture
of young stars from dwarf companions.  Well known examples of
interacting and/or merging early type galaxies are Centaurus~A=NGC~5128
and Fornax~A=NGC~1316, which are strong radio sources and indeed have
produced 1+2 SNIa in the last century.  Moreover, the Virgo Cluster
radiogalaxy NGC 4753, which is a parent galaxy to one of the SNIa in
Evans' sample (SN1983G), had given birth to another SNIa (SN1965I) less
than two decades before, thus confirming its high SNIa rate.
Also in favor of this hypothesis is the fact that the observed SNIa
rates are higher in late type galaxies than in early type galaxies
(about a factor of 10 between Ellipticals and Sd galaxies) once the
rates are normalized to the galaxy H or K band luminosities, and,
therefore, to their masses in stars, rather than to their B-band
magnitudes (Della Valle \& Livio 1994, Panagia 2000, Della Valle
\etal~2003).  This fact confirms the connection between recent stellar
populations and type Ia supernovae, and provides a natural explanation
of why among early type galaxies are SNIa rates higher in
radiogalaxies, without requiring the concurrence of any new and/or
additional process to enhance the SNIa rates.
In addition, we note that recently Howell (2001) (see also Ivanov,
Hamuy and Pinto 2000 ) has found convincing evidence that SNIa
properties correlate with the host galaxy types.  In particular, out of
a total sample of 67 SNIa selected from Branch, Fisher and Nugent
(1999), Li \etal~(2001), Barbon \etal~(1999) and Phillips et al. 1999
compilations, Howell found that: 
{\it i)} Underluminous SNIa are twice more common in early type galaxies
than in late type galaxies (17 {\it vs} 8),  
{\it ii)} Overluminous SNIa are much more common in late type
galaxies than in early type galaxies (15 {\it vs} 2), and  
{\it iii)} Overluminous SNIa in early type galaxies have been found in
radio-loud galaxies only (2).  
These facts can easily be understood in terms of a picture in which
most early type galaxies have to ``scrap the bottom of the barrel" for
SNIa progenitors and only in those galaxies that have undergone recent
interaction/merging episodes may one find the overluminous SNIa that
are believed to be produced in younger systems from the most massive
SNIa progenitors (\eg Tutukov \& Yungelson 1994, 1996, Ruiz-Lapuente,
Burkert \& Canal 1995). On the other hand, late type galaxies, thanks
to their intrinsically active star formation, and regardless of whether
they are interacting/merging with other galaxies, have an abundant and
steady supply of SNIa progenitors of all appropriate masses and,
therefore, can have a higher production of overluminous SNIa. A
corollary is that the average absolute luminosity of SNIa at maximum
light is higher for late type galaxies than early type galaxies, as
observed (Della Valle \& Panagia 1992, Hamuy et al. 1995, Hamuy
\etal~2000).  It is also expected that a number of SNIa, but not all,
be produced in the direct vicinities of radio jets and/or radio lobes
because at the early stages of the interaction those are the locations
where star formation may be highest. Clearly, this scenario can easily
account for all different aspects of SNIa properties in galaxies of
different morphological types and different radio activity.

Finally, we note that in our scenario the occurrence of a few type
II-Ib/c SNe in early type galaxies is expected at the early stages of
the same star formation burst that provides a steady supply of SNIa
events.  Since core collapse SNe are produced by progenitors more
massive than 8M$_\odot$, whose lifetime is shorter than about 30 Myrs,
only for a small fraction of the time between subsequent
interaction/capture episodes (approximately 30Myrs/1Gyr=3\%) would one
expect to see such SNe in early type galaxies.  Since in galaxies with
active star formation the rate of core collapse SNe is about 4 times
higher than that of SNIa, we estimate that one should expect 
core collapse SNe to occur at an average rate of about 0.03$\times4\simeq$ 12\%
that of SNIa events.  This, however, is still not the probability of
detecting core-collapse SNe as compared to SNIa because we have to
consider  that core-collapse SNe are generally fainter than SNIa, and
that the former may go undetected because of strong dust extinction
(such as present in the dust lane of Cen~A).  Neglecting dust
extinction, but folding in the facts that SNII/SNIbc have a median peak
luminosity that is  about two magnitudes fainter than SNIa and that as
many as  $\sim$25\% of the core-collapse SNe may be as bright as SNIa
(Richardson \etal~2002), we estimate that the overall probability of
{\it actually detecting}  core-collapse SNe in a magnitude
limited-sample is about 4\% of that of SNIa events.  Such a small
percentage, which may be further reduced by extinction, can explain why
no SNII/SNIbc have been detected in radio-loud elliptical galaxies so
far. Nevertheless, we predict  that  increasing the sample of
radio-loud elliptical galaxies searched appreciably, one will
eventually detect core-collapse SNe in such early type galaxies. 
\bigskip

We thank Paolo Padovani for useful discussions on radio galaxies, and
an anonymous referee for valuable comments. MDV is grateful to the
STScI for the friendly hospitality and creative atmosphere.

\newpage

\centerline{\bf FIGURE CAPTIONS}

\figcaption{The distribution of early type galaxies in Evans
\etal~(1989) sample as a function of their absolute B-band magnitudes. 
The solid line is the distribution of the entire sample, the shaded
histogram is the distribution of radio-loud galaxies, and the dashed
histogram is the general luminosity function of elliptical galaxies,
adapted from Muriel \etal~(1995).
\label{fig:galaxies}}

\vskip .3in

\begin{thebibliography}{}
\bibitem[]{} Baade, W., \& Minkowski, R. 1954, ApJ, 119, 206
\bibitem[]{} Balick, B., \& Heckman, T.M. 1982, ARA\&A, 20, 431
\bibitem[]{} Barbon, R. Buondi, V., Cappellaro, E., \& Turatto,
	M. 1999, A\&AS, 139, 531
\bibitem[]{} Bicknell, G.V. 2002, New AR, 46, 365
\bibitem[]{} Branch, D, Fisher, A., \& Nugent, P. 1993, AJ, 106, 2383
\bibitem[]{} Canizares, C.R., Fabbiano, G., \& Trinchieri, G. 1987, ApJ, 312,
	503
\bibitem[]{} Capetti, A. 2002, ApJ, 574, L25
\bibitem[]{} Cappellaro, E., Evans, R., \& Turatto, M. 1999, A\&A, 351, 459
\bibitem[]{} Della Valle, M., \& Livio, M. 1994, ApJ,423, L31
\bibitem[]{} Della Valle, M., \& Panagia, N. 1992, AJ, 104, 696
\bibitem[]{} Evans, R, van den Bergh, S., \& Clure, R.D. 1989, ApJ, 345,
	752
\bibitem[]{} Geldzahler, B. J., \& Fomalont, E. B. 1984, AJ, 89, 1650
\bibitem[]{} Gioia, I.M., Gregorini, L., \& Vettolani, P. 1981, A\&A, 96, 58
\bibitem[]{} Hamuy, M., Phillips, M.M., Maza, J., Suntzeff, N.B.,
	Schommer, R.A.,  \& Aviles, R. 1995, AJ, 109,1
\bibitem[]{}Hamuy, M., Trager, S. C., Pinto, P.A., Phillips, M. M.,
	Schommer, R. A., Ivanov, V., \& Suntzeff, N.B. 2000, AJ, 120, 1479
\bibitem[]{} Heckman, T.M., Smith, E.P., Baum, S.A., van Breugel,
	W.J.M., Miley, G.K., Illingworth, G.D., Bothun, G.D., \& Balick, B.
	1986, ApJ, 311, 526
\bibitem[]{} Howell, D.A. 2001, ApJ, 554, L193
\bibitem[]{} Ivanov, V.D., Hamuy, M., \& Pinto, P.A. 2000, ApJ, 542, 588
\bibitem[]{} Ledlow, M.J., \& Owen, F.N. 1996, AJ, 112, 9
\bibitem[]{} Li, W.-D., Filippenko, A.,V., Treffers, R.R., Riess, A.G.,
	Hu, J., \& Qiu, Y. 2001, ApJ, 546, 734
\bibitem[]{} Livio, M., Riess, A., \& Sparks, W. 2002, ApJ, 571, L99
\bibitem[]{} Madau, P., Della Valle, M., \& Panagia, N. 1998, MNRAS, 297, L17
\bibitem[]{} Mannucci, F., Maiolino, R., Cresci, G., Della Valle, M., Vanzi, L.,
             Ghinassi, F., Ivanov, V.D., Nagar, N.M., Alonso-Herrero, A. 2003, A\&A, in press
             (astro-ph/0302323)
\bibitem[]{} Muriel, H. Nicotra, M. A., \& Lambas, D.G. 1995, AJ, 110, 1032
\bibitem[]{} Pain, R. \etal~2002, astro-ph/0205476, ApJ, in press
\bibitem[]{} Panagia, N. 2000, in {\it ``Experimental Physics of
	Gravitational Waves"}, eds. G. Calamai, M. Mazzoni,  R. Stanga \& F.
	Vetrano,  (World Scientific -- Singapore) p. 107-119
\bibitem[]{} Phillips, M.M., Lira, P., Suntzeff, N.B., Schommer, R.A.,
	Hamuy, M., \& Maza, J. 1999, AJ, 118, 1766
\bibitem[]{} Prialnik, T., \& Kovetz,A. 1995, ApJ, 445, 789 
\bibitem[]{} Renzini, A., Ciotti, L., D'Ercole, A., \& Pellegrini, S.
	1993, ApJ, 419, 52
\bibitem[]{} Ruiz-Lapuente, P., Burkert, A., \& Canal, R. 1995, ApJ, 447, L69
\bibitem[]{} Richardson, D., Branch, D., Casebeer, D., Millard, J.,
	Thomas, R.C., \& Baron, E. 2002, AJ, 123, 745
\bibitem[]{} Sadler, E.M., Jenkins, C.R., \& Kotanyi C.G. 1989, MNRAS, 240, 591
\bibitem[]{} Sandage, A., \& Tammann, G.A. 1981, A Revised Shapley-Ames
	Catalog of Bright Galaxies (Washington: Carnegie Institution) 
\bibitem[]{} Schechter, E.M. 1976, ApJ, 203, 297
\bibitem[]{} Tutukov, A.V., \& Yungelson, L. 1994, MNRAS, 268, 871
\bibitem[]{} Tutukov, A.V., \& Yungelson, L. 1996, MNRAS, 280, 1035
\bibitem[]{} van den Bergh, S., Clure, R.D., \& Evans, R. 1987, ApJ, 323, 44
\bibitem[]{} van den Bergh, S., Li, W., Filippenko, A.V. 2002, PASP, 114, 820
\bibitem[]{} Weiler, K.W., Panagia, N., Montes, M., Sramek, R.A. 2002, ARA\&A, 40, 387
\bibitem[]{} Whitmore, B.C., Zhang, Q. Leitherer, C., Fall, S. M.,
	Schweizer, F., \& Miller, B.W. 1999, AJ, 118, 1551
\bibitem[]{} Urry, C.M., \& Padovani, P. 1995, PASP, 107, 803
\end{thebibliography}
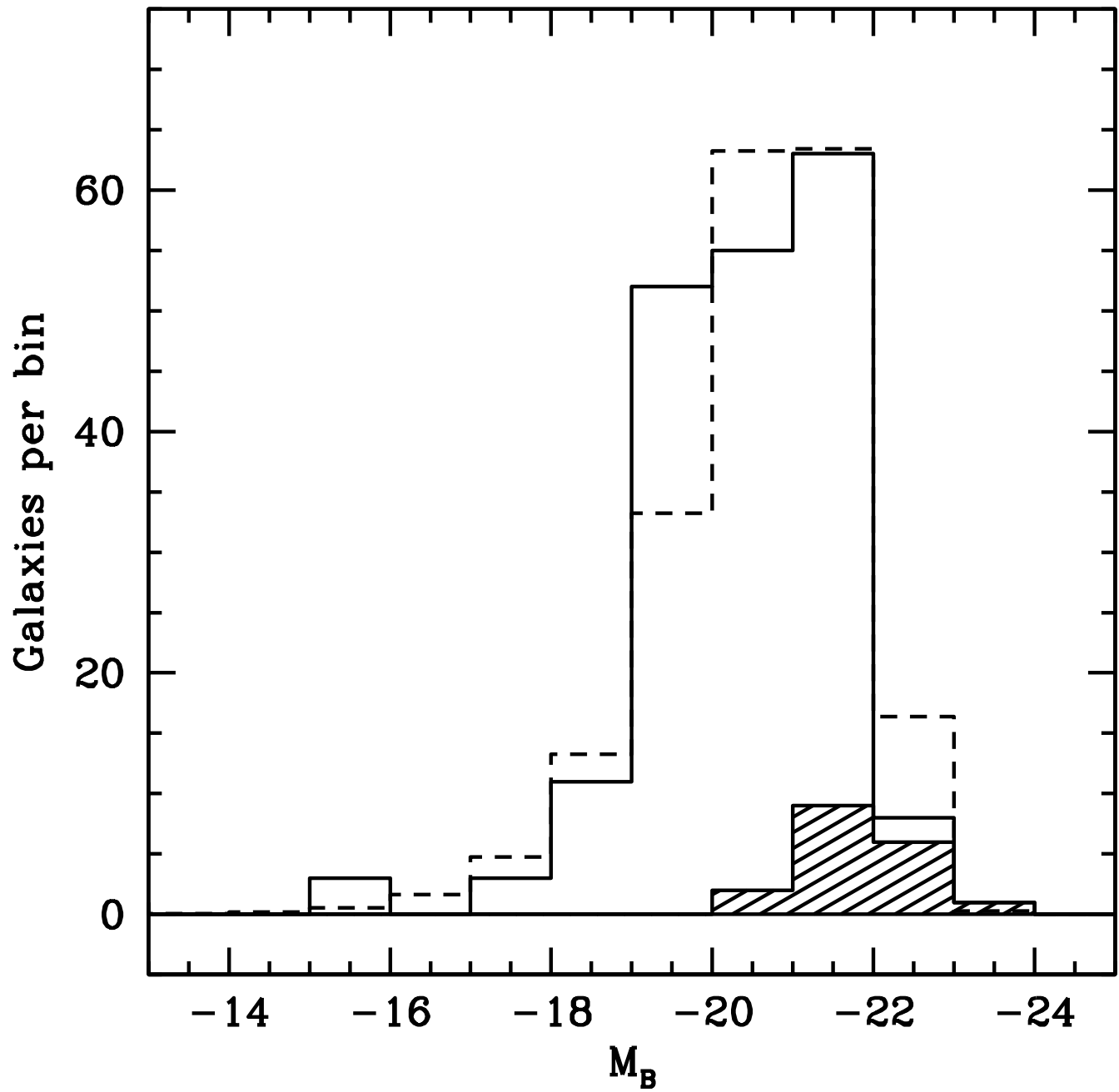
\end{document}